\documentclass{article}

\usepackage{booktabs} 
\usepackage{graphicx} 
\usepackage{subfigure} 
\usepackage{amssymb} 
\usepackage{wrapfig} 
\usepackage{xspace}
\usepackage[latin2]{inputenc}
\usepackage{indentfirst}
\usepackage{enumerate}
\usepackage{times}

\begin{document}

\title{Hadron production measurement from NA61/SHINE}

\begin{center}
{\Large Hadron production measurement from NA61/SHINE}
\end{center}

\begin{center}
{A.Korzenev\footnote{Presented at EPS-HEP\,2013 in Stockholm on July 19, 2013.}, on behalf of the NA61/SHINE collaboration}

{DPNC, University of Geneva, Switzerland}

{korzenev@mail.cern.ch}
\end{center}

\begin{abstract}
New results of NA61/SHINE on determination of
charged hadron yields in proton-carbon interactions are presented.
They aim to improve predictions of the neutrino flux
in the T2K experiment. The data were recorded using
a secondary-proton beam of 31 GeV/$c$ momentum from CERN SPS
which impinges on a graphite target. 
To determine the inclusive production cross section for charged 
pions, kaons and protons the thin ($0.04\, \lambda_I$) target was exploited. 
Results of this measurement are used in the T2K beam simulation program
to reweight hadron yields in the interaction vertex.
At the same time,
NA61/SHINE results obtained with the T2K replica target ($1.9\, \lambda_I$)
allow to constrain hadron yields at the surface of the target. It
would correspond to the constraint up to 90\% of the neutrino flux,
thus reducing significantly a model dependence of the neutrino beam prediction.
All measured spectra are compared to predictions of hadron production models. 

In addition a status of the analysis of data collected by NA61/SHINE
for the NuMI target (Fermilab) is reviewed. 
These data will be used further in neutrino beam calculations for 
the MINERvA, MINOS(+) and LBNE experiments.
\end{abstract}





A precise prediction of the expected neutrino flux is required for the T2K experiment
\cite{review,T2K_flux_paper}.
It is used to calculate a neutrino cross section at the near detector,
while at the far detector it provides an estimate of
the expected signal for the study of neutrino oscillations. 
Prediction of the neutrino flux  is constrained by using dedicated hadron
production measurements. These measurements were therefore performed 
in the NA61/SHINE experiment \cite{proposal_NA61} at CERN.
We refer the reader to \cite{NA49} for the description of main detector components,
software, calibration and analysis methods which were basically inherited from 
the NA49 experiment.

The cross section measurement using a secondary-proton beam of 31 GeV/$c$ momentum
from CERN\,SPS scattered off a thin graphite target
(0.04 $\lambda_{I}$) have been performed by NA61 in years 2007 and 2009. 
The first NA61 physics papers were devoted to the 
production of $\pi^\pm$ and K$^+$ \cite{Abgrall:2011ae,Abgrall:2011ts}. 
For this analysis  data collected in 2007 have been used.
By now these results were integrated to the T2K beam simulation program 
to constrain the production of hadrons in the primary interaction of 
the beam protons in the target \cite{Abe:2012gx,Abe:2011sj}.

\begin{figure}
\centering
\includegraphics[width=0.3\textwidth]{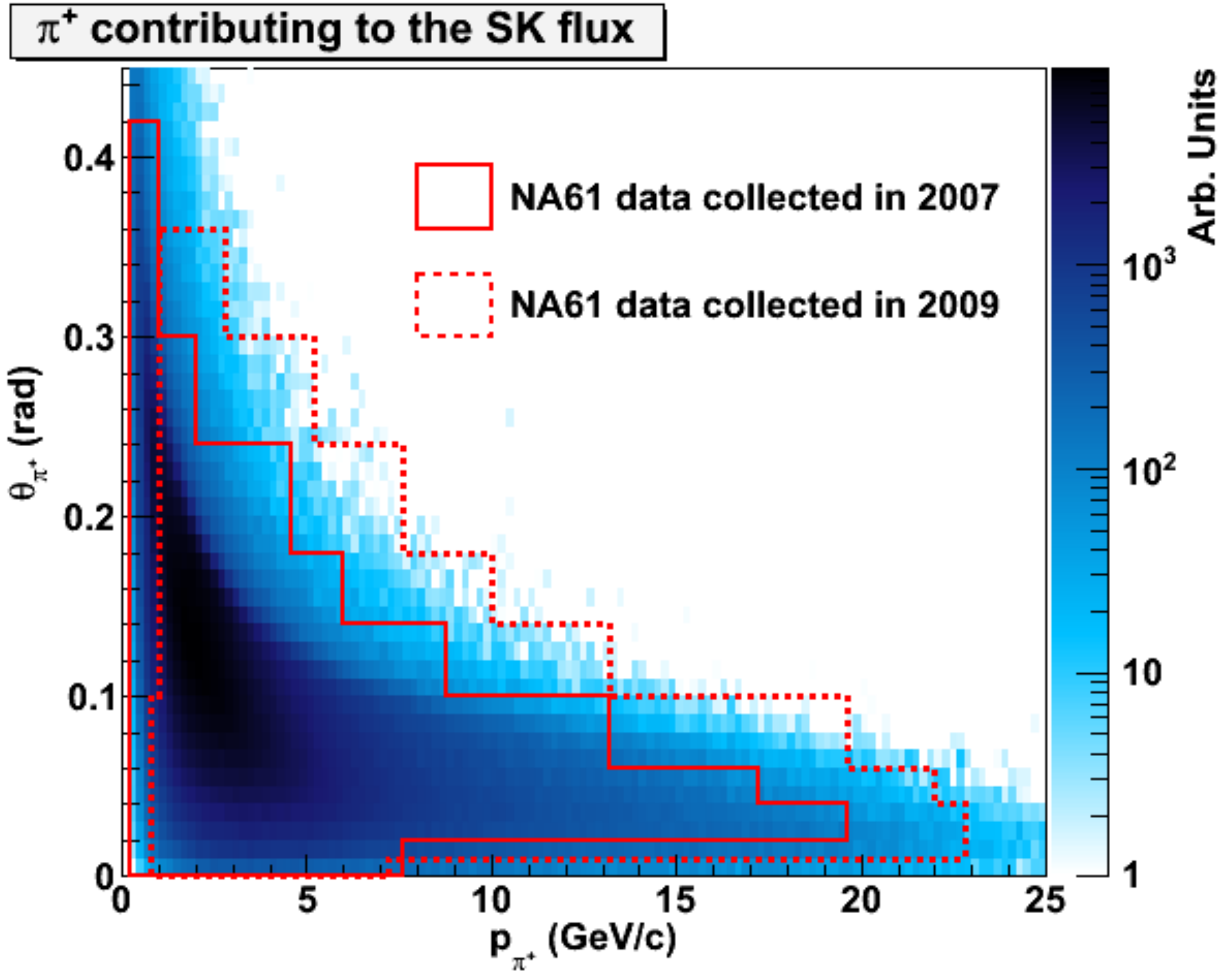}
\includegraphics[width=0.3\textwidth]{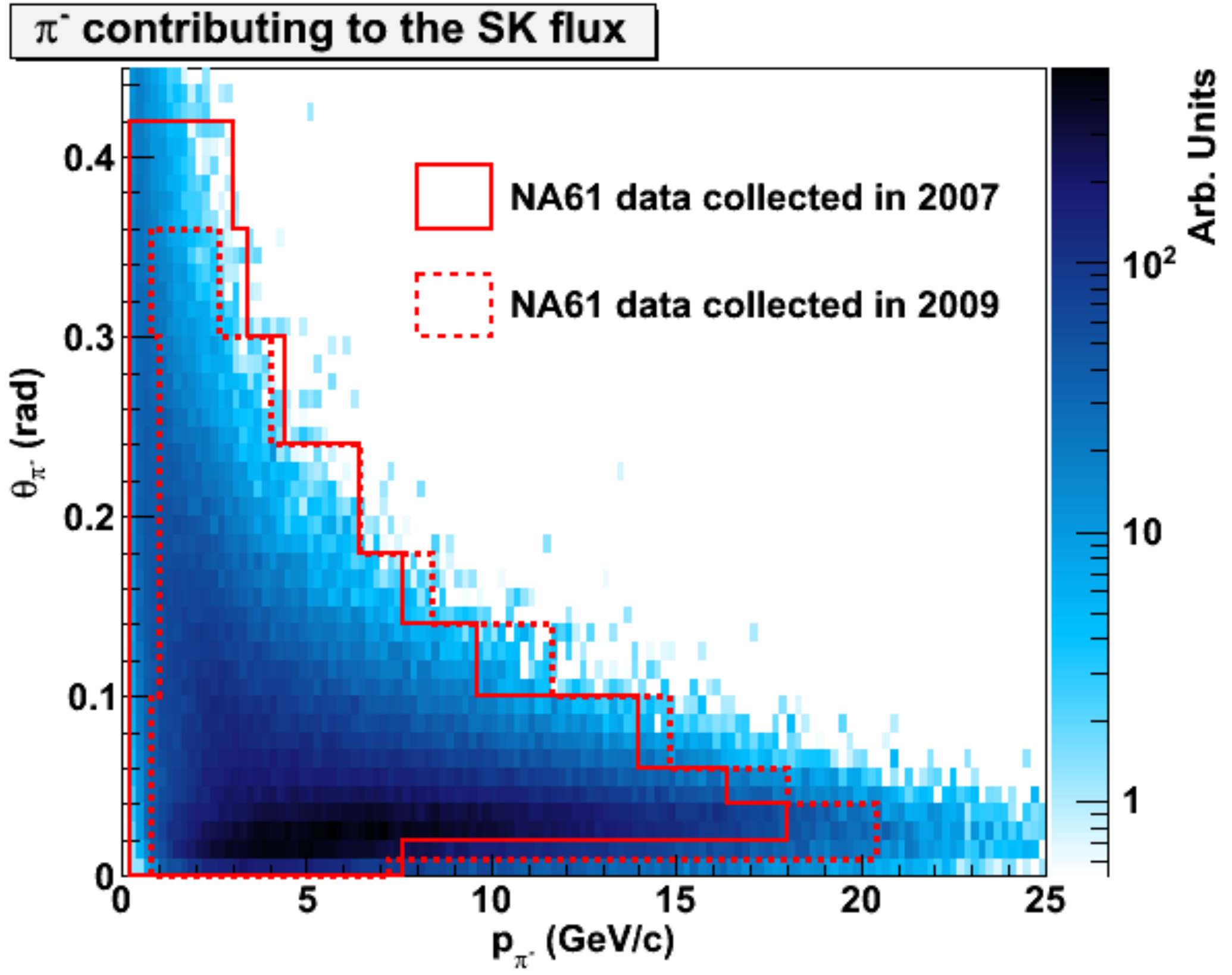}
\includegraphics[width=0.3\textwidth]{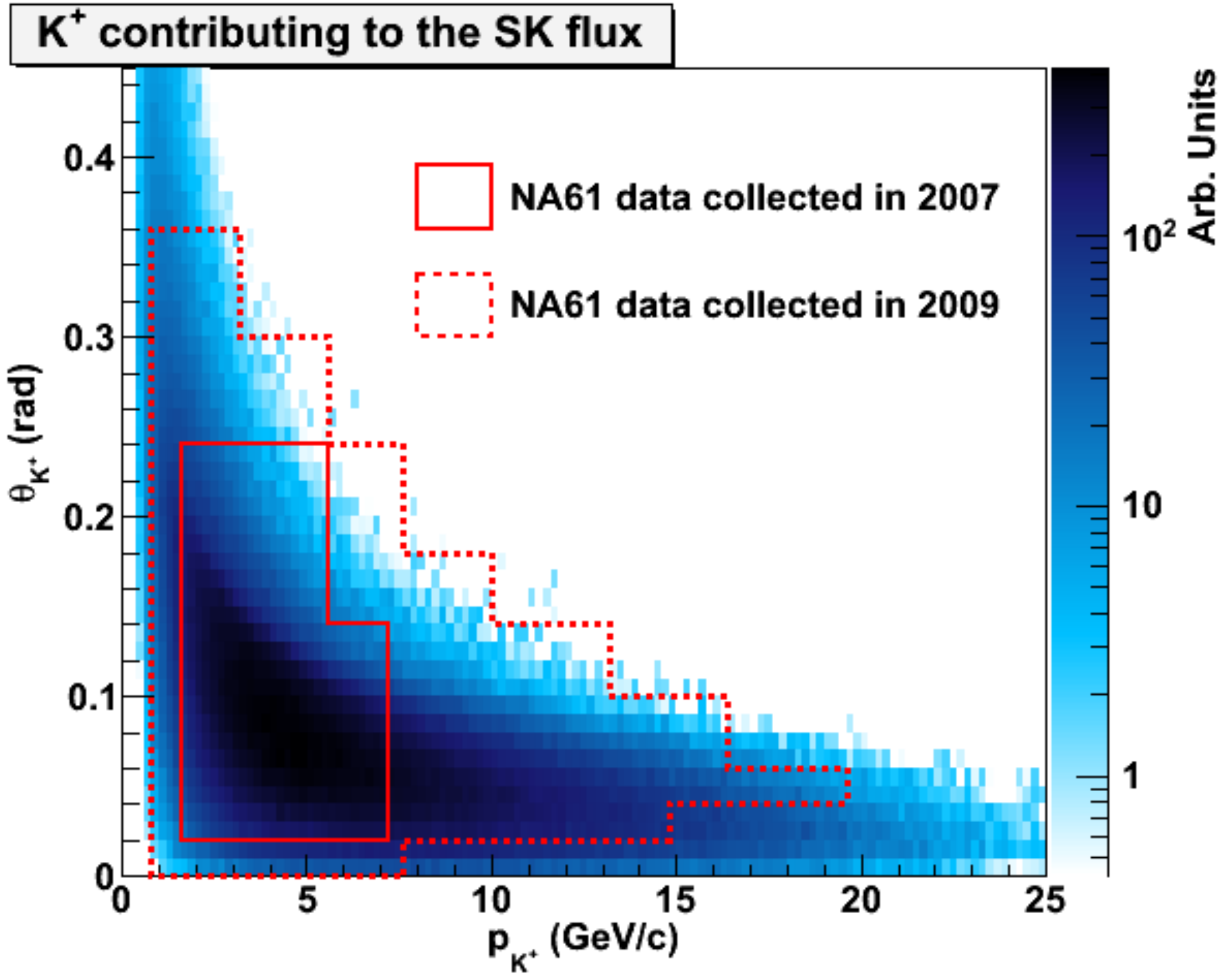}
\includegraphics[width=0.3\textwidth]{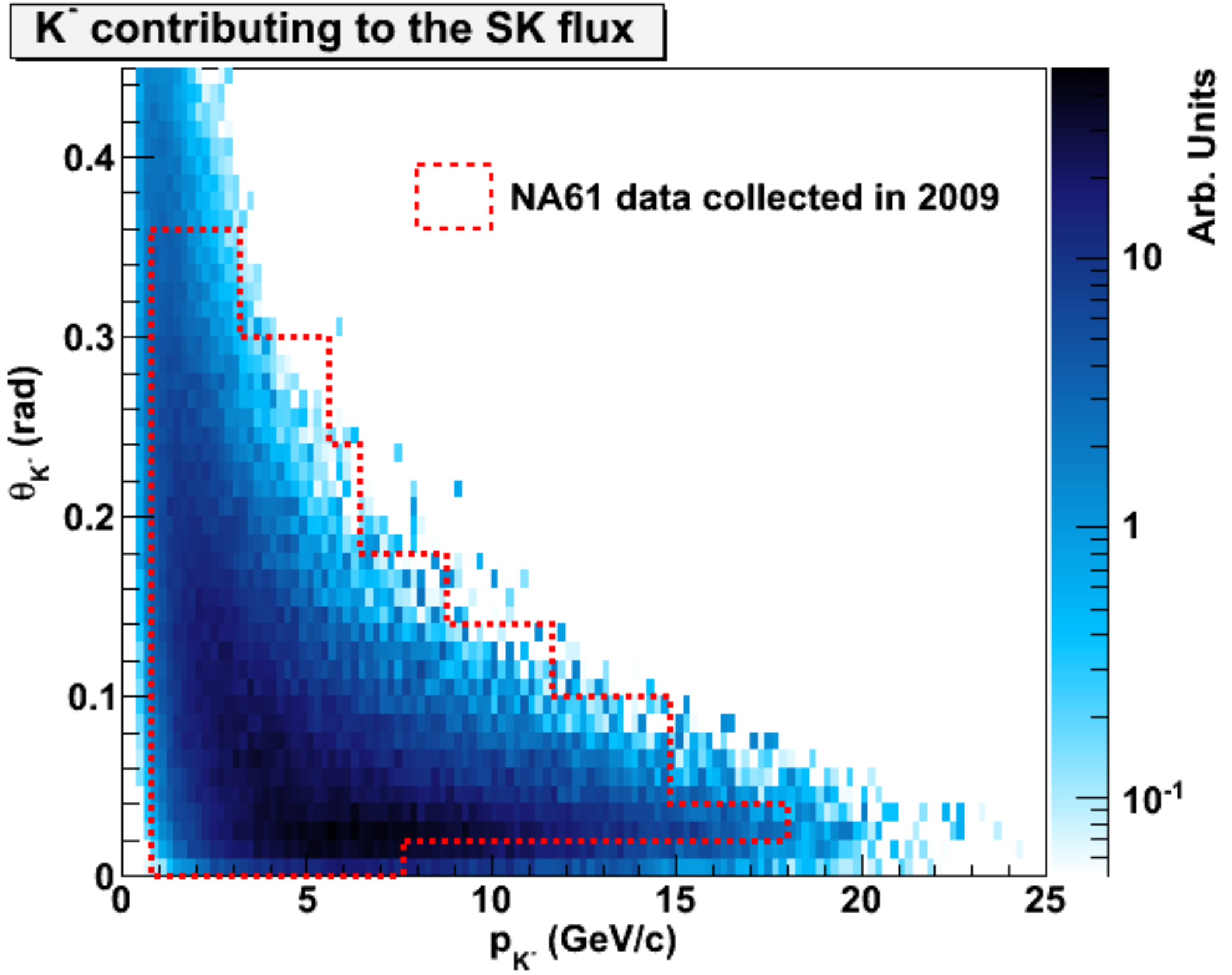}
\includegraphics[width=0.3\textwidth,height=0.242\textwidth]{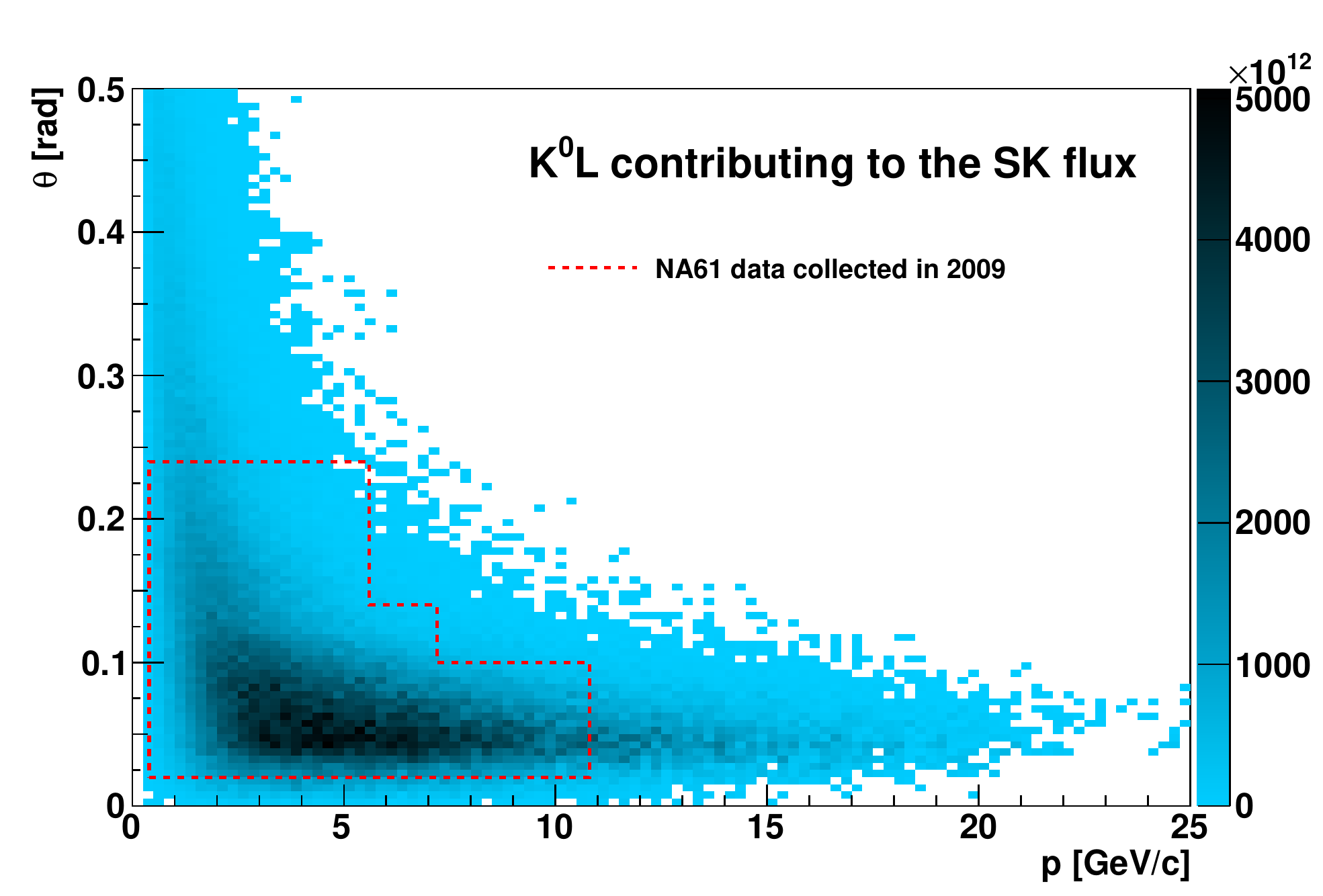}
\includegraphics[width=0.3\textwidth]{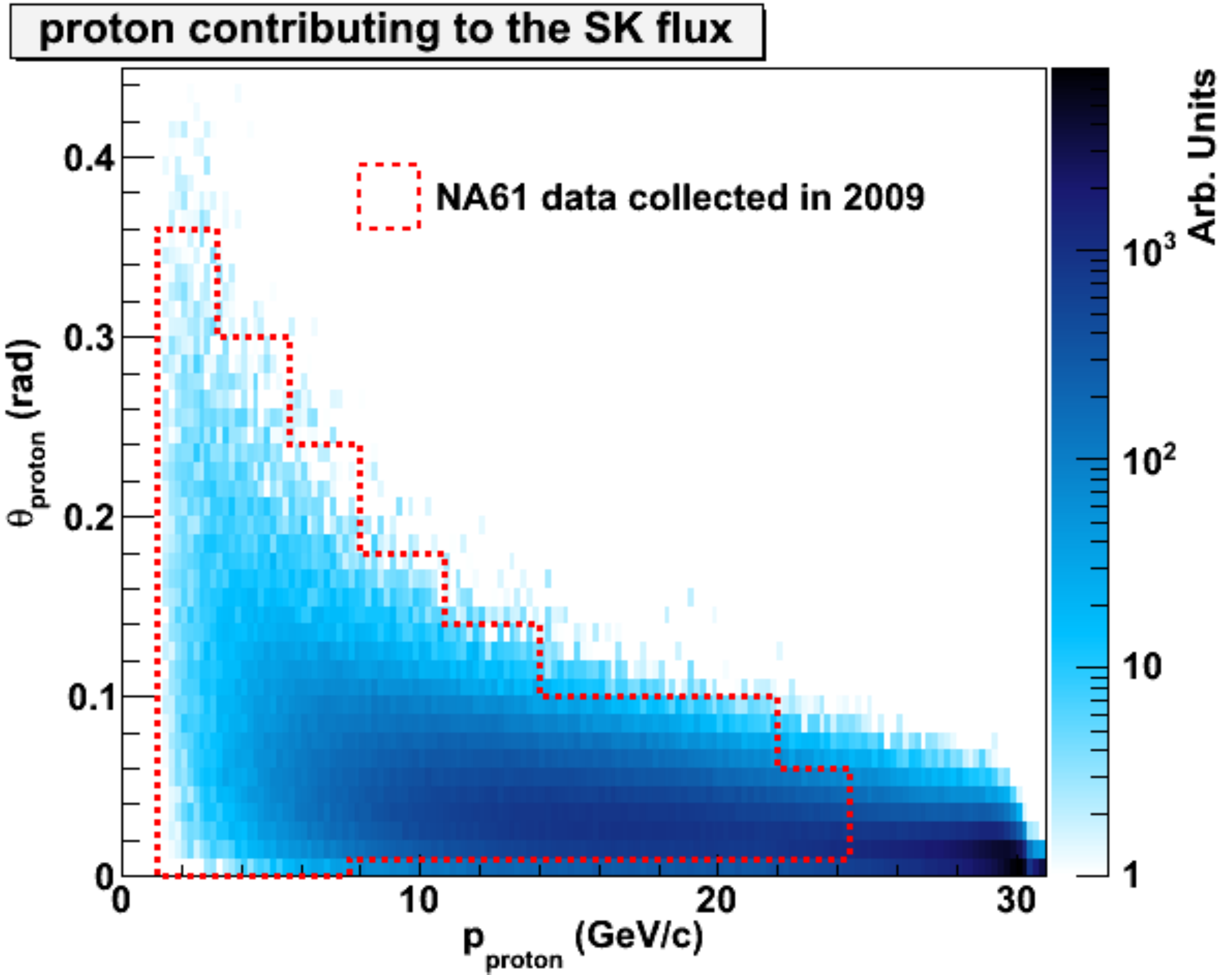}
\caption{ The phase space of $\pi^+$, $\pi^-$, K$^+$, K$^-$, K$^0_L$
  and protons contributing to the predicted neutrino flux at SK 
  in the ``positive'' focusing configuration \cite{T2K_flux_paper}, 
  and the regions covered by new 2009 data ({\it dashed line}) and by
  previously published NA61/SHINE
  measurements ({\it solid line}) \cite{Abgrall:2011ae,Abgrall:2011ts}.}
\label{fig:na61_coverage}
\end{figure}

Although pilot data 2007 covered a significant part of the relevant hadron production 
phase space of T2K \cite{T2K_flux_paper} the statistical uncertainty is quite large.
In the year 2008 important changes have been introduced to the experimental 
setup of NA61: new trigger logic, TPC read-out and DAQ upgrade,
additional sections of ToF wall, new beam-telescope detectors.
As a consequence of these upgrades the number of events recorded in 2009 and 2010
for about a same period of time have been increased by an order of magnitude
as compared to the 2007 data. This larger sample allows simultaneous extraction 
of yields of $\pi^{\pm}$, K$^{\pm}$, K$^0_s$ and protons.
Furthermore the phase space of NA61 has been increased (see Fig.\,\ref{fig:na61_coverage}).
Additional sections of ToF improve the coverage 
at high $\theta$.
In the forward direction one profits from the use of the Gap TPC detector. 
It plays a key role in the analysis of forward produced particles.
The coverage of this kinematic domain is important for the muon monitor 
measurements in T2K \cite{Abe:2011ks}. 
Statistics collected in 2007 and 2009 with the thin target is 
$0.7$ and $5.4$ millions of triggers, respectively.

An essential fraction of the neutrino flux arises
from secondary re-interactions as long targets are used in T2K 
\cite{T2K_flux_paper,Abgrall:2012pp}. 
The lack of direct data, hence the use of sparse data sets, to cover these 
contributions limits the achievable precision on the flux prediction. 
Therefore measurements with a full-size replica of the T2K target (1.9 $\lambda_{I}$) 
have been performed by NA61 in years 2007, 2009 and 2010. 
A total of $0.2$, $4$ and $10$ millions of triggers 
have been recorded on tape, respectively. 
Analysis of these data will allow to reduce the systematic uncertainties of 
neutrino flux due to the treatment of secondary re-interactions in the target.

\subsubsection*{Normalization and production cross section}

For the normalization of hadron spectra and the calculation of the production 
cross section we use a procedure described in \cite{Abgrall:2011ae,NA49}.
The idea is to
measure an interaction probability for cases when the graphite target
was inserted and removed. Using these values one calculates the so-called
``trigger'' cross section which, in turn, is an input for the analysis of
``physics'' cross sections.
Due to upgrade of the spectrometer and higher beam intensity 
in 2009 the procedure was slightly modified as compared to the analysis 
of the 2007 data.


\begin{wrapfigure}{r}{0.46\textwidth}
\centering
\vspace*{-0.3cm}
\includegraphics[width=0.45\textwidth]{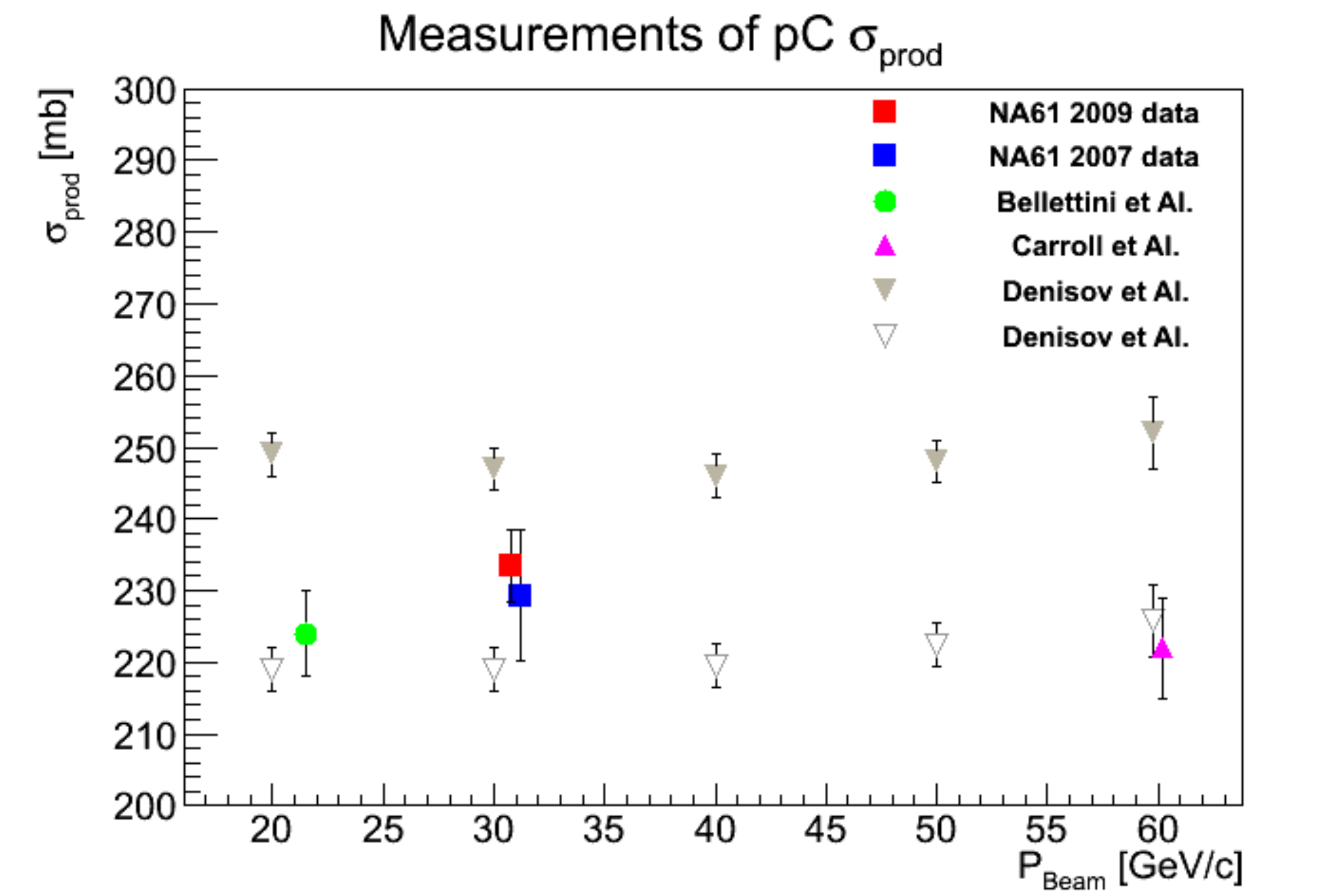}
\caption{A comparison of the production cross section to
  the previously published results obtained at different beam momenta
  \cite{Bellettini,Carroll,Denisov}.
}
\label{Davide}

\vspace*{0.1cm}
\end{wrapfigure}

In the 2009 campaign the beam trigger ran simultaneously with physics triggers.
The ability to apply the event-by-event selection improved the systematics significantly.
To account for the high rate the beam trigger was prescaled by a factor of 100.

As a result of the analysis of the 2009 data we obtain the inelastic cross section
\begin{eqnarray}
\sigma_{inel} = 261.3 \pm 2.8(\mbox{stat})  \pm 2.2 (\mbox{model}) \pm 1.0 (\mbox{trig})~{\rm mb} 
\nonumber 
\end{eqnarray}
which comprises all processes due to strong \mbox{p + C} 
interactions excluding the coherent nuclear elastic scattering. 
By subtracting from $\sigma_{inel}$ the cross section of quasielastic interactions, 
which amounts to $27.8 \pm 2.2$ (stat) mb, one determines the production cross section 
\begin{eqnarray}
\sigma_{prod} = 233.5 \pm 2.8(\mbox{stat}) \pm 4.2(\mbox{model}) \pm 1.0 (\mbox{trig})~{\rm mb}.\nonumber
\end{eqnarray}
It is used further to normalize hadron cross sections to be able to compare to 
MC models.
All model-dependent corrections were estimated with GEANT4.9.5~\cite{GEANT4,GEANT4bis}
using FTF\_BIC physics list.
A comparison to the previously published results is presented in Fig.\,\ref{Davide}.

The total uncertainty of $\sigma_{prod}$ is 5.1 mb, almost a factor of two
smaller than the one obtained with the 2007 data. 
The dominant contribution to the uncertainty comes from the physics model 
used to recalculate the production cross section from the ``trigger'' cross section.

\subsubsection*{Measurement of charged hadron cross sections for T2K}

Depending on the momentum interval and the particle type, different
analysis techniques were tested for the analysis of data 2007 \cite{Abgrall:2011ae}. 
More than 90\% of primary negatively charged particles produced 
at this energy are $\pi^-$. Thus the analysis of
$\pi^-$ spectra can be done without additional particle
identification (so called $h^-$ approach).
For other species of particles the identification is mandatory.
In particular for the region of momenta $p>1$ GeV/$c$, 
which is the most crucial for T2K neutrino kinematics,
a combined analysis of time-of-flight (ToF) measurements and energy loss, $dE/dx$,
measurements in TPC was used.

\begin{figure*}[t]
\centering
\includegraphics[width=0.45\textwidth]{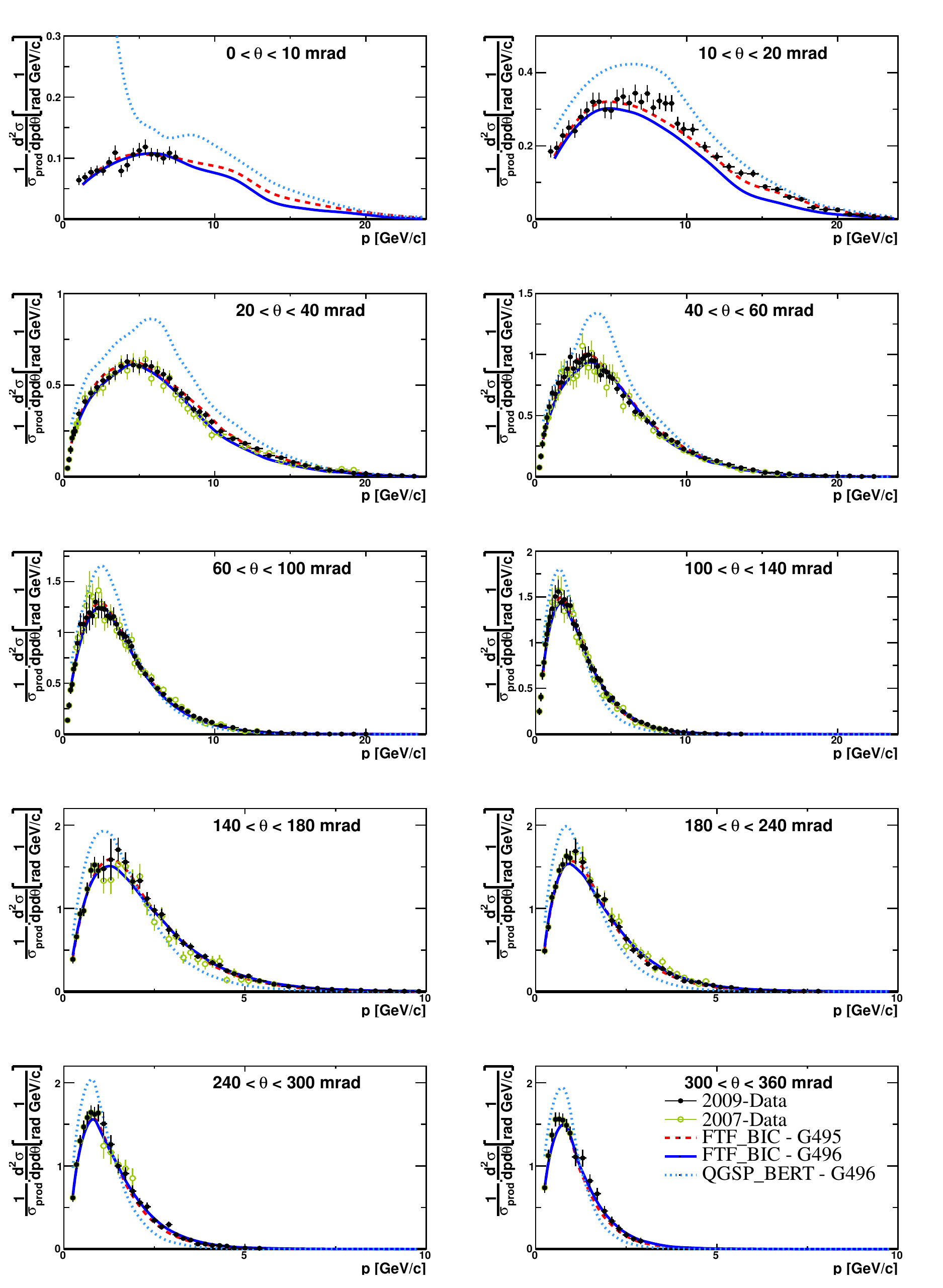}
\hspace{1cm}
\includegraphics[width=0.45\textwidth]{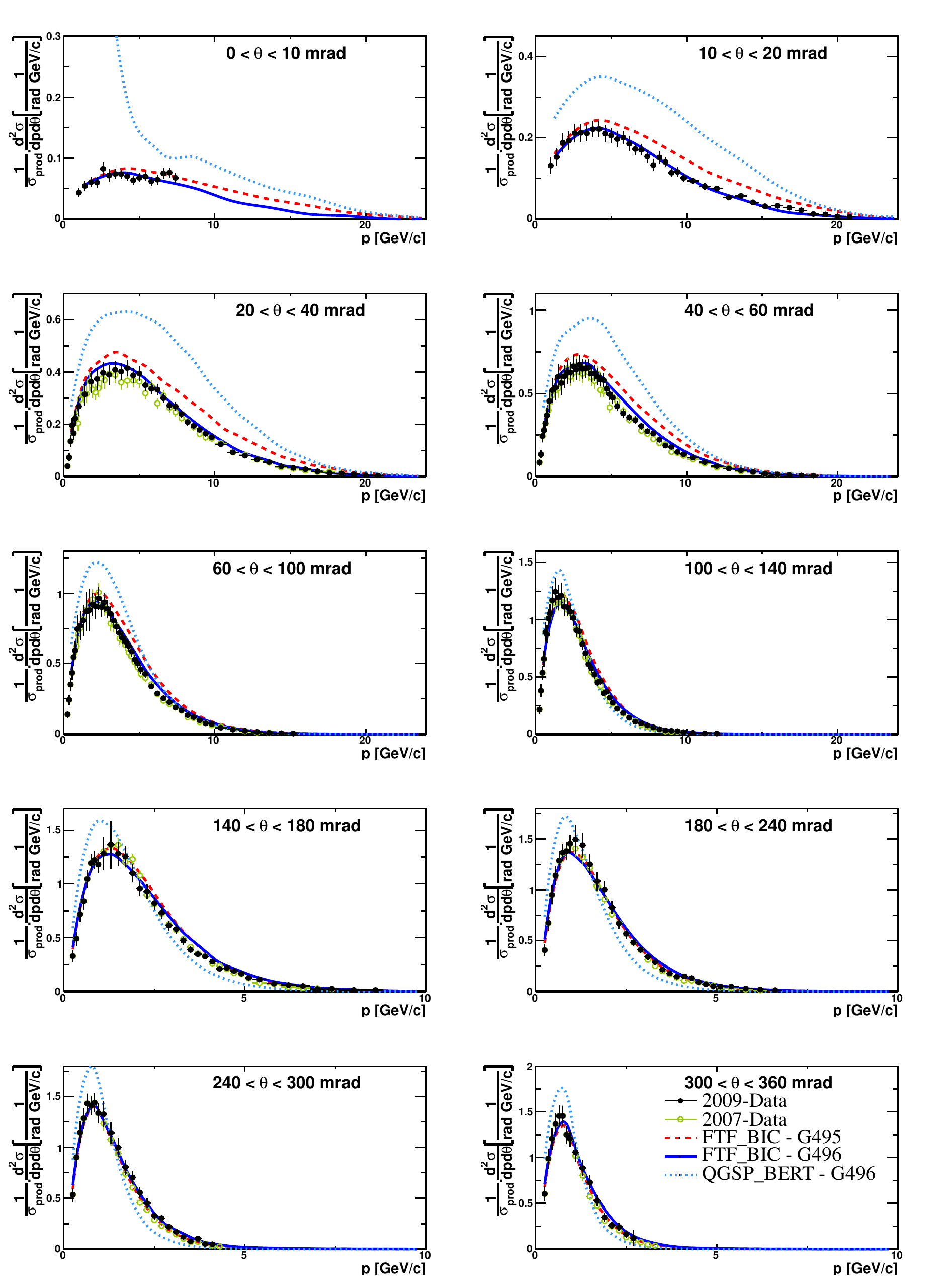}
\caption{Laboratory momentum distributions of the $\pi^+$ and $\pi^-$ multiplicities 
produced in p-C interactions at 31 GeV/$c$ in different intervals of polar angle $\theta$.
Error bars indicate statistical uncertainty.
Data points are overlapped by various model predictions \cite{GEANT4,GEANT4bis}.
}
\label{NA61_pions}
\end{figure*}

Raw particle yields have been corrected step-by-step using 
the NA61 Monte Carlo simulation program with VENUS \cite{VENUS} 
for primary interactions and
a GEANT\,3-based part for tracking of secondary particles through the detector.
The following effects have been accounted for: 
geometrical acceptance of the spectrometer; 
efficiency of the reconstruction chain; decays before reaching the ToF wall; 
ToF detection efficiency; pions coming from $\Lambda$ and K$^0_s$ decays 
(feed-down correction).

Spectra of $\pi^{+}$ and $\pi^{-}$ obtained with data 2007 and 2009
are presented  in Fig.\,\ref{NA61_pions}. 
For data 2007 three analysis techniques were applied: 
$dE/dx$ method for momenta smaller than 1 GeV/c; combined ToF-$dE/dx$
method for momenta larger than 1 GeV/c; $h^-$ method for $\pi^-$ 
cross section.
Spectra were compared in overlapping regions to check their consistency.
Complementary domains were combined to reach maximum acceptance.
Due to large statistics in 2009 it was decided to use only 
the ToF-$dE/dx$ method since it provides high selection purity 
of the sample and low dependence on a MC model.

\begin{figure*}[t]
\centering
\includegraphics[width=0.45\textwidth]{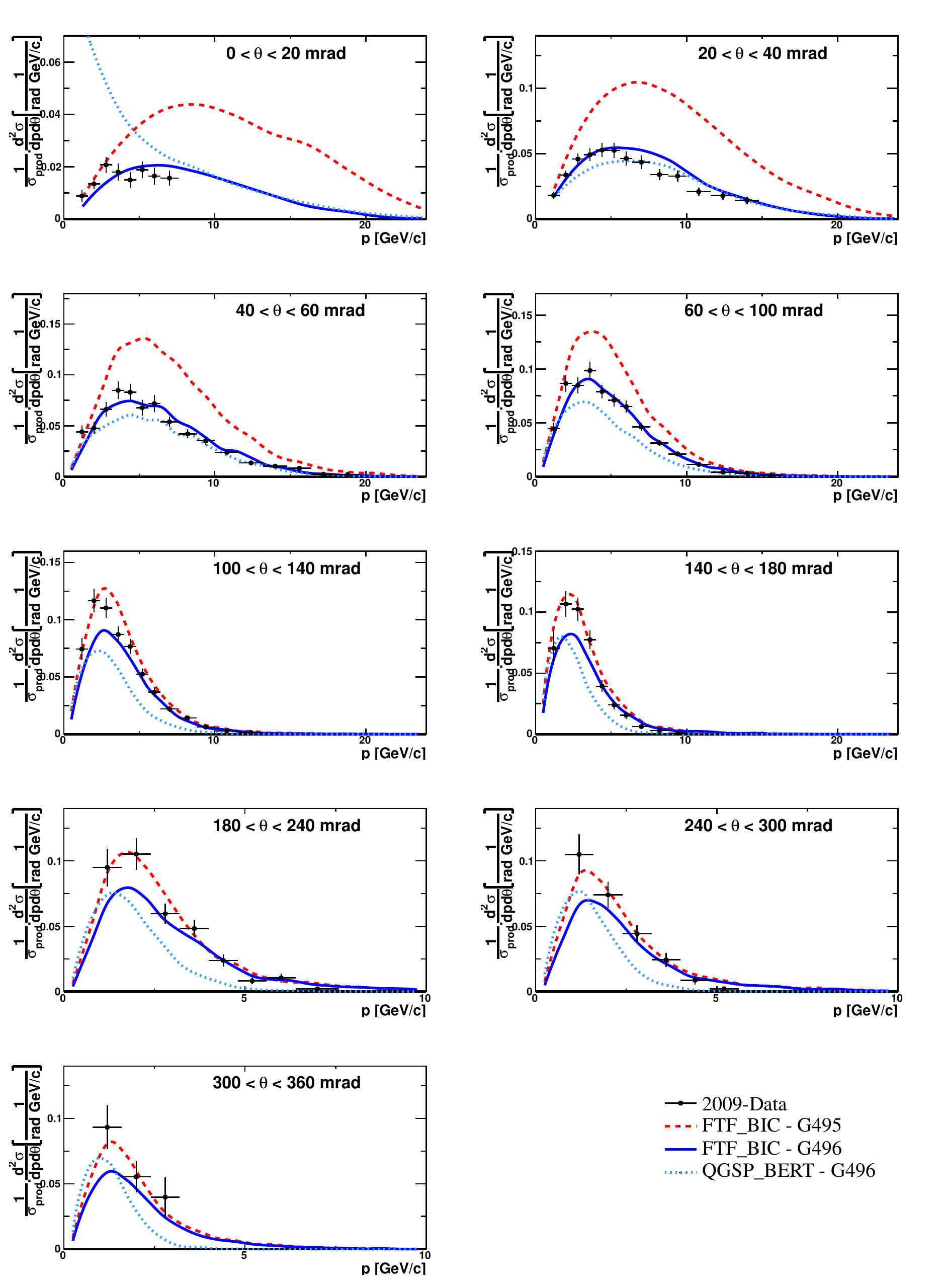}
\hspace{1cm}
\includegraphics[width=0.45\textwidth]{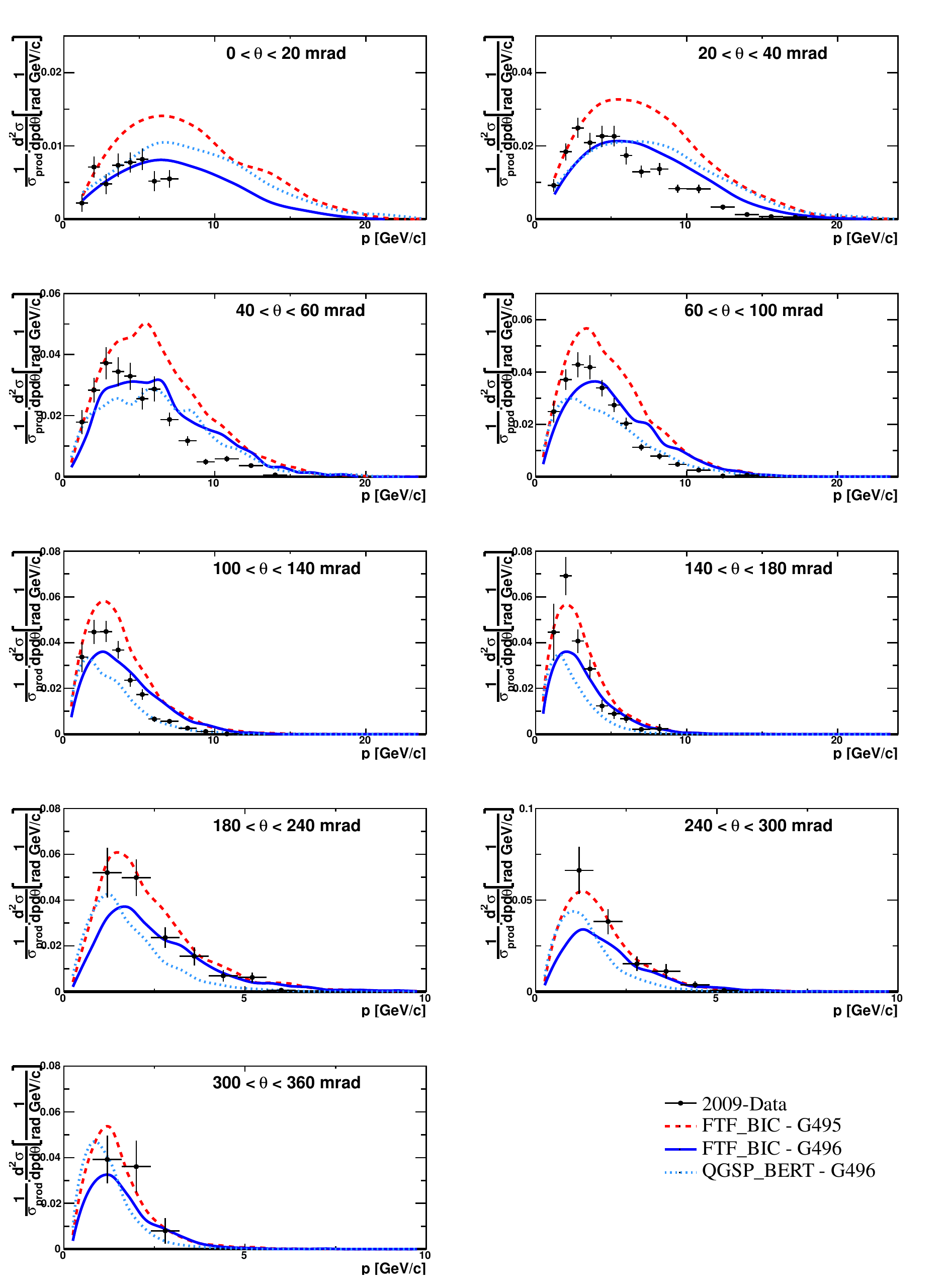}
\caption{Laboratory momentum distributions of the K$^+$ and K$^-$ multiplicities 
produced in p-C interactions at 31 GeV/$c$ in different intervals of polar angle $\theta$.
Error bars indicate statistical uncertainty.
Data points are overlapped by various model predictions \cite{GEANT4,GEANT4bis}. 
}
\label{NA61_kaons}

\end{figure*}

The analysis of K$^\pm$ is more complicated 
due to their small fraction in the overall sample \cite{Abgrall:2011ts}. 
For instance the K$^+$ signal vanishes over the predominant pion one 
at the low momentum range while at higher momenta protons dominate.
Analysis technique is basically similar to the ToF--$dE/dx$ one used for pions.
In general statistical error of the K$^+$ spectra with data 2007 
is by a factor of 3 larger than the systematic one \cite{Abgrall:2011ts} 
and only two intervals in $\theta$ were considered.
Data collected in 2009 improved the precision strongly.
In Fig.\ref{NA61_kaons} the differential cross section of kaons 
normalized to the production cross section is shown. 
Larger statistics of 2009 allowed to split the phase space into 
9 intervals in $\theta$.
Graphs are overlaid with predictions from several recommended                      
GEANT4 physics lists \cite{GEANT4,GEANT4bis}.

\begin{figure*}[t]
\centering
\begin{minipage}[b]{0.49\linewidth}
\includegraphics[width=0.99\textwidth]{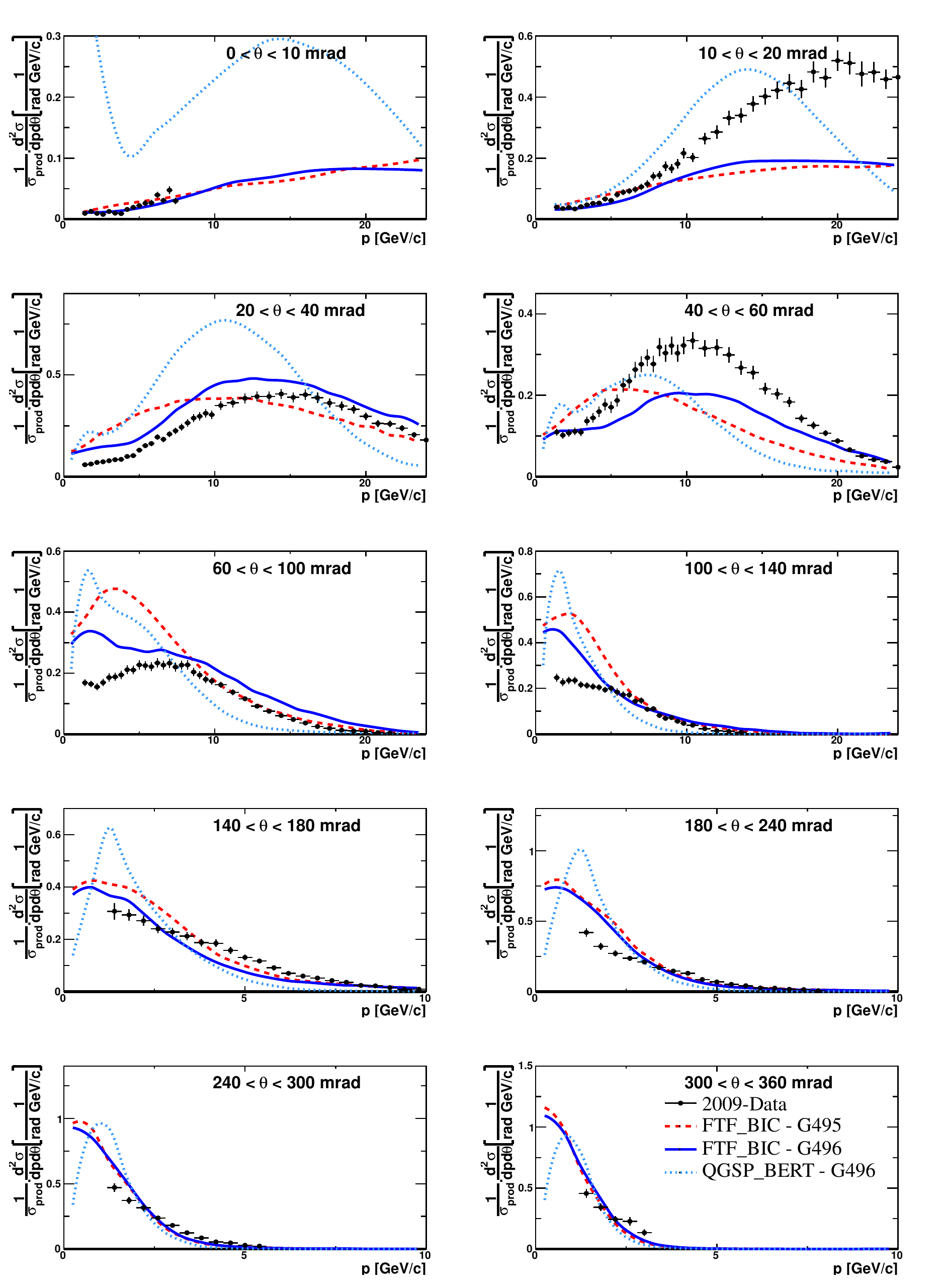}
\caption{
Laboratory momentum distributions of the proton multiplicities 
produced in p-C interactions at 31 GeV/$c$ in different intervals of polar angle $\theta$.
Error bars indicate statistical uncertainty.
Data points are overlapped by various model predictions \cite{GEANT4,GEANT4bis}.
}\label{NA61_protons}
\end{minipage}
\hfill
\begin{minipage}[b]{0.49\linewidth}
\centering
\includegraphics[width=0.99\textwidth]{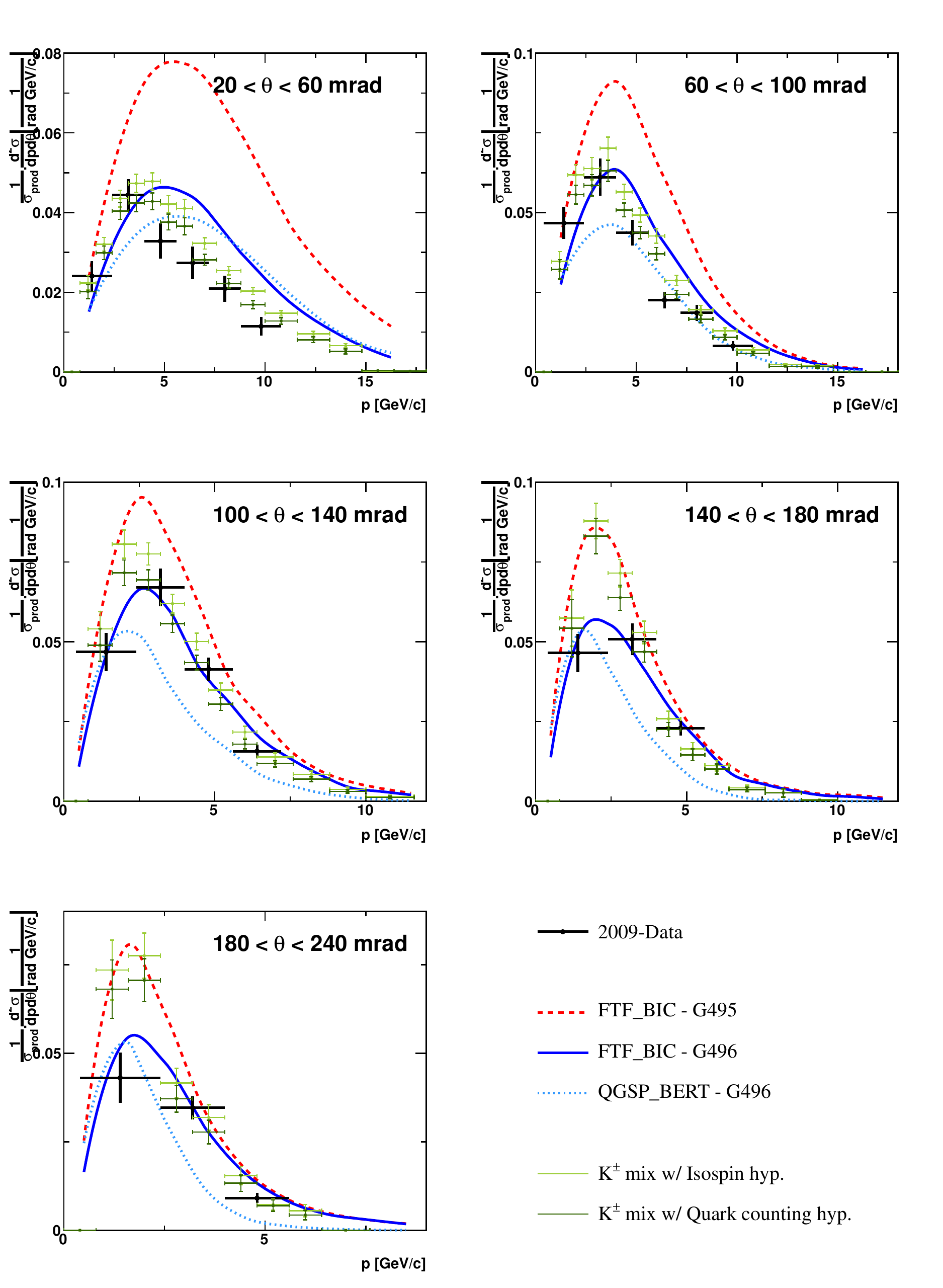}
\caption{
Laboratory momentum distributions of the $K^0_S$ multiplicities 
produced in p-C interactions at 31 GeV/$c$ in different intervals of polar angle $\theta$.
Error bars indicate statistical uncertainty.
Data points are overlapped by various model predictions 
\cite{GEANT4,GEANT4bis}.
}
\label{NA61_K0S}
\end{minipage}
\end{figure*}

Distribution of proton multiplicities obtained with data 2009
as a function of momentum in different intervals of $\theta$ 
is shown in Fig.\,\ref{NA61_protons}.
Comparison to GEANT4 models shows that none of them 
can satisfactory describe the data.

\subsubsection*{Cross section of neutral particles $K_S^0$ and $\Lambda$}

Understanding of the neutral strange particle production
in NA61/SHINE
is of great interest for two reasons. 
First, it allows to decrease the systematic uncertainty associated with 
the charged particle production, namely pions \cite{Abgrall:2011ae} and protons. 
Second, the measurement of the $K^0_S$ production will improve 
our knowledge of the $\nu_e$ flux at the T2K experiment coming from 
the three body decay of the $K^0_L \to \pi^0 e^\pm \nu_e (\overline \nu_e)$ 
\cite{T2K_flux_paper}.

The technique used for the analysis of the 2009 data have been
already tested on the 2007 data \cite{K0_2007}.
We reconstruct $K_S^0$ and $\Lambda$  in so-called $V^0$ mode:
decay into two charged particle of opposite signs
$K^0_S \to \pi^+ + \pi^-$ and
$\Lambda \to p \phantom{^+} + \pi^-$.
Particle yields have been extracted in the analysis of invariant mass spectra
applying corresponding mass hypotheses for daughter tracks.
The main background sources were associated with
converted photons and the combinatorial background which is mainly 
due to particles produced in the primary interaction.
Selection cuts have been applied on the following variables:
the number of clusters for daughter tracks, 
the impact parameter of the $V^0$ track in the primary vertex, 
the distance of closest approach between daughter tracks, 
the distance between primary and decay vertices, 
the cosine of the angle between the trajectory of $V^0$ and 
daughter particle in the center-of-mass system of $V^0$, 
the energy loss in TPC by daughter tracks,
and finally 
$V^0$ satisfying the hypothesis of $K_S^0$ were removed from 
the analysis of $\Lambda$ and vice versa.
Multiplicities of $K_S^0$ are shown in Fig.\,\ref{NA61_K0S} together with 
a prediction of several models.
Analysis of $\Lambda$ is in progress.



Being able to measure simultaneously K$^+$, K$^-$ and K$^0_S$ we can 
provide a test for several hypotheses predicting a relative yield 
of charged and neutral kaons. These hypotheses derived from the isospin
symmetry and/or some basic assumption on parton distributions in nucleon \cite{BMPT}.
%
%
The comparison of the measured differential multiplicity of K$^0_S$ 
to the prediction obtained with the charged kaons is shown in Fig.\,\ref{NA61_K0S}.
A reasonable agreement is observed for prediction of 'isospin' and 'quark-counting' 
hypotheses presented in the figure.
However more certain statement 
would require a higher statistical precision for K$^0_S$.


\subsubsection*{Analysis of the T2K replica-target data}

First physics results from the analysis of the replica-target 
data taken in 2007 have been published \cite{Abgrall:2012pp}. 
A dedicated reconstruction method has been developed to provide results 
in a form that is of direct interest for T2K.  Yields of
positively charged pions are reconstructed at the surface of the T2K
replica target in bins of the laboratory momentum and polar angle
as a function of the longitudinal position along the target. 
 By parametrizing hadron yields on a surface of the target
one predicts up to 90\% of the flux for both $\nu_{\mu}$ 
and $\nu_e$ components while only 60\% of neutrinos are 
coming from decay of particles produced in the primary interaction.
Two methods (constraint of 
hadroproduction data at primary interaction and on a target surface) 
are consistent within their uncertainties achieved on statistics of pilot run 2007.
The ultimate precision will come from the analysis of the replica-target data 
2009 and 2010.

\subsubsection*{Data taking for the Fermilab neutrino beam}

Following discussion initiated at NUFACT\,2011 
a group from 10 US institutions   expressed their interest in possibility 
of collecting  data relevant for NuMI experiments (MINERvA, NOvA, MINOS+), 
Booster experiments (MiniBooNE, MicroBooNE) and LBNE \cite{Schmitz:2012}.
An important pilot run 
with a proton beam of 120 GeV and a thin graphite target 
took place in July 2012. It resulted to 3.5 millions of recoded events.
An experience with these pilot data gives a basis to estimate
an amount of efforts needed to complete the Fermilab's neutrino program in NA61. 
A complete data taking is expected after the 2013/2014 shutdown of 
the CERN accelerator complex.



\end{document}